\documentclass[aps,prl,twocolumn,preprintnumbers,amsmath,amssymb,superscriptaddress,10pt]{revtex4}
\usepackage{graphicx}
\usepackage{bm}
\usepackage[usenames,dvipsnames]{color}

\pdfoutput=1 

\begin{document}

\title{Spin-orbit coupling in a graphene bilayer and in graphite}

\author{F. Guinea}
\affiliation{Instituto de Ciencia de Materiales de Madrid, CSIC, Sor
Juana In\'es de la Cruz 3, E28049 Madrid, Spain}

\begin{abstract}
The intrinsic spin-orbit interactions in bilayer graphene and in
graphite are studied, using a tight binding model, and an
intraatomic $\vec{\bf L}\vec{\bf S}$ coupling. The spin-orbit
interactions in bilayer graphene and graphite are larger, by about
one order of magnitude, than the interactions in single layer
graphene, due to the mixing of $\pi$ and $\sigma$ bands by
interlayer hopping. Their value is in the range $0.1 - 1$K. The
spin-orbit coupling opens a gap in bilayer graphene, and it also
gives rise to two edge modes. The spin-orbit couplings are largest,
$\sim 1-4$K, in orthorhombic graphite, which does not have a center
of inversion.
\end{abstract}

\pacs{73.20.-r; 73.20.Hb; 73.23.-b; 73.43.-f}

\maketitle


\section{Introduction}
The isolation and control of the number of carriers in single and
few layer graphene flakes\cite{Netal04t,Netal05t} has lead to a
large research activity exploring all aspects of these
materials\cite{NGPNG09}. Among others, the application of graphene
to spintronic
devices\cite{HGNSB06t,CCF07,NG07,HJPJW07t,TTVJJvW08t,WPLCWSK08t,Petal09}
and to spin qubits\cite{TBLB07,FTL09,WYMK09} is being intensively
studied. The understanding of these devices requires a knowledge of
the electronic spin-orbit interaction. In principle, this
interaction turns single layer graphene into a topological
insulator\cite{KM05}, which shows a bulk gap and edge states at all
boundaries. The magnitude of the spin-orbit coupling in single layer
graphene has been studied\cite{HGB06,HMING06t,YYQZF07,GKEAF09}. The
calculated couplings are small, typically below 0.1K. The observed
spin relaxation\cite{TTVJJvW08t,Jetal09} suggests the existence of
stronger mechanisms which lead to the precession of the electron
spins, like impurities or lattice
deformations\cite{NG09,HGB09,EKGF09}.

Bilayer graphene is interesting because, among other properties, a
gap can be induced by electrostatic means, leading to new ways for
the confinement of electrons\cite{MF06}. The spin-orbit interactions
which exist in single layer graphene modulate the gap of a graphene
bilayer\cite{GM09}. The unit cell of bilayer graphene contains four
carbon atoms, and there are more possible spin-orbit couplings than
in single layer graphene.

We analyze in the following the intrinsic and extrinsic spin-orbit
couplings in bilayer graphene, using a tight binding model, and
describing the relativistic effects responsible for the spin-orbit
interaction by a $\vec{{\bf L}}\vec{{ \bf S}}$ intraatomic coupling.
We use the similarities between the electronic bands of a graphene
bilayer and the bands of three dimensional graphite with Bernal
stacking to generalize the results to the latter.

\section{The model}.
We describe the electronic bands of a graphene bilayer using a tight
binding model, with four orbitals, the $2s$ and the three $2p$
orbitals,  per carbon atom. We consider hoppings between nearest
neighbors in the same plane, and nearest neighbors and next nearest
neighbors between adjacent layers, see\cite{CLM09}. The couplings
between each pair of atoms is parametrized by four hoppings, $V_{ss}
, V_{sp} , V_{pp \pi}$ and $V_{pp \sigma}$. The model includes also
two intraatomic levels, $\epsilon_s$ and $\epsilon_p$, and the
intraatomic spin-orbit coupling
\begin{align}
{\cal H}_{so} &\equiv \Delta_{so} \sum_i \vec{\bf L}_i \vec{\bf S}_i
\end{align}
 The parameters used
to describe the $\pi$ bands of graphite\cite{M57,SW58}, $\gamma_0 ,
\gamma_1 , \gamma_2 , \gamma_3 , \gamma_4 , \gamma_5$ and $\Delta$,
can be derived from this set of parameters. We neglect the
difference between different hoppings between atoms which are next
nearest neighbors in adjacent layers, which are responsible for the
difference between the parameters $\gamma_3$ and $\gamma_4$. We also
set the difference in onsite energies between the two inequivalent
atoms, $\Delta$ to zero. The parameters $\gamma_2$ and $\gamma_5$
are related to hoppings between next nearest neighbor layers, and
they do not play a role in the description of the bilayer. The total
number of parameters is 15, although, without loss of generality, we
set $\epsilon_p = 0$. We do not consider hoppings and spin orbit
interactions which include $d$ levels, although they can contribute
to the total magnitude of the spin-orbit
couplings\cite{MY62,GKEAF09}. The effects mediated by $d$ orbitals
do not change the order of magnitude of the couplings in single
layer graphene, and their contribution to interlayer effects should
be small.

The main contribution to the effective spin-orbit at the Fermi level
due to the interlayer coupling is due to the hoppings between $p$
orbitals in next nearest neighbor atoms in different layers. This
interaction gives rise to the parameters $\gamma_3$ and $\gamma_4$
in the parametrization of the bands in graphite. For simplicity, we
will neglect couplings between $s$ and $p$ orbitals in neighboring
layers. The non zero hoppings used in this work are listed in
Table~\ref{hoppings}.

\begin{table}
\begin{tabular}{||c|c||} \hline \hline
$\epsilon_s$ &-7.3  \\ \hline  $t^0_{ss}$ & 2.66 \\
\hline $t^0_{sp}$ & 4.98  \\ \hline  $t^0_{pp \sigma}$  &2.66
\\ \hline $t^0_{pp \pi}$ &-6.38 \\ \hline  $t^1_{pp \pi}$ &0.4 \\ \hline
$t^2_{pp \sigma}$ &0.4 \\ \hline  $t^2_{pp \pi}$ &-0.4  \\ \hline
 $\Delta_{so}$ &0.02 \\ \hline \hline
\end{tabular}
\caption{Non zero tight binding parameters, in eV, used in the
model. The hoppings are taken from\cite{TL87,TS91}, and the
spin-orbit coupling from\cite{SCR00}. Superindices 0,1, and 2
correspond to atoms in the same layer, nearest neighbors in
different layers, and next nearest neighbors in different layers.}
\label{hoppings}
\end{table}

The hamiltonian can be written as a $32 \times 32$ matrix for each
lattice wavevector. We define an effective hamiltonian acting on the
$\pi$, or $p_z$, orbitals, by projecting out the rest of the
orbitals:
\begin{align}
{\cal H}_{\pi}^{eff} &\equiv {\cal H}_{\pi} + {\cal H}_{\pi \sigma}
\left( \omega - {\cal H}_{\sigma \sigma} \right)^{-1} {\cal
H}_{\sigma \pi} \label{heff}
\end{align}
We isolate the effect of the spin-orbit coupling by defining:
\begin{align}
{\cal H}_{\pi}^{so} \left( \vec{\bf k} \right) &\equiv {\cal
H}_{\pi}^{eff} ( \Delta_{so} ) - {\cal H}_{\pi}^{eff} ( \Delta_{so}=
0 )
\end{align}
Note that ${\cal H}_{\pi}^{so}$ depends on the energy, $\omega$.

\begin{figure}
\includegraphics[width=5cm]{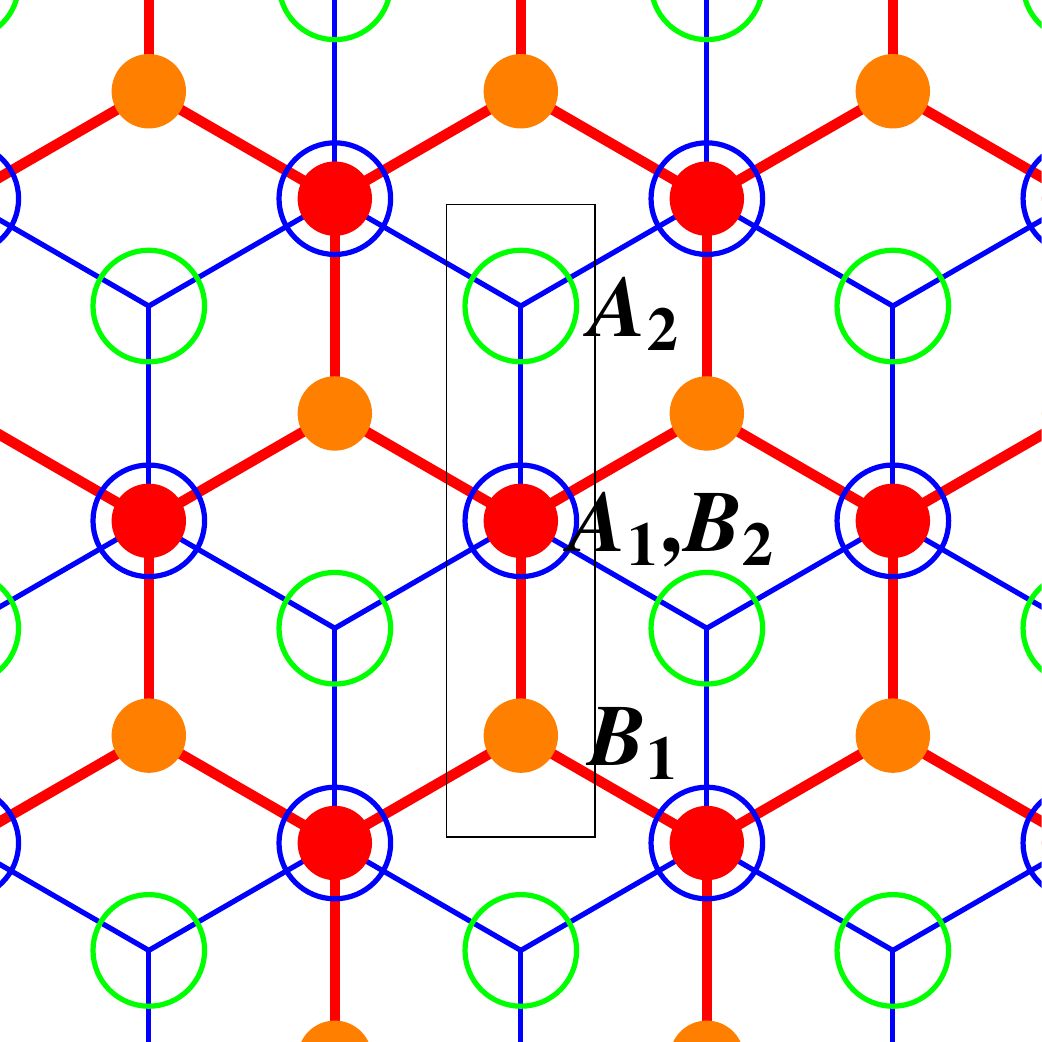}
\caption[fig]{(Color online).  Unit cell of a graphene bilayer.
Labels A and B define the two sublattices in each layer, while
subscripts 1 and 2 define the layers.} \label{bilayer_lattice_so}
\end{figure}

We analyze ${\cal H}_{\pi}^{so}$ at the $K$ and $K'$ points. The two
matrices have a total of 16 entries, which can be labeled by
specifying the sublattice, layer, spin, and valley. We define
operators which modify each of these degrees of freedom using the
Pauli matrices $\hat{\sigma} , \hat{\mu} ,  \hat{s}$, and
$\hat{\tau}$. The unit cell is described in
Fig.~\ref{bilayer_lattice_so}.

The hamiltonian has inversion and time reversal symmetry, and it is
also invariant under rotations by $120^\circ$. These symmetries are
defined by the operators:
\begin{align}
{\cal I} &\equiv \sigma_x \mu_x \tau_x \nonumber \\
{\cal T} &\equiv i s_y \tau_x {\cal K} \nonumber \\
{\cal C}_{120^\circ} &\equiv \left( - \frac{1}{2}  + i
\frac{\sqrt{3}}{2} s_z \right) \times \left( - \frac{1}{2} - i
\frac{\sqrt{3}}{2} \tau_z \mu_z \right) \times \nonumber \\ &\times
\left( - \frac{1}{2} + i \frac{\sqrt{3}}{2} \tau_z \sigma_z \right)
\end{align}
where ${\cal K}$ is complex conjugation.

The possible spin dependent terms which respect these symmetries
were listed in\cite{DD65}, in connection with the equivalent problem
of three dimensional Bernal graphite (see below). In the notation
described above, they can be written as
\begin{align}
{\cal H}_{\pi}^{so} &= \lambda_1 \sigma_z \tau_z s_z + \lambda_2
\mu_z \tau_z s_z +  \lambda_3 \mu_z \left( \sigma_y s_x - \tau_z
\sigma_x s_y \right) + \nonumber \\ &+ \lambda_4 \sigma_z \left(
\mu_y s_x + \tau_z \mu_x s_y \right) \label{hamilso}
\end{align}
The first term describes the intrinsic spin-orbit coupling in single
layer graphene. The other three, which involve the matrices $\mu_i$,
are specific to bilayer graphene. The term proportional to
$\lambda_3$ can be viewed as a Rashba coupling with opposite signs
in the two layers.

\section{Results}.
\subsection{Bilayer graphene}.
\begin{figure}
\includegraphics[width=8cm]{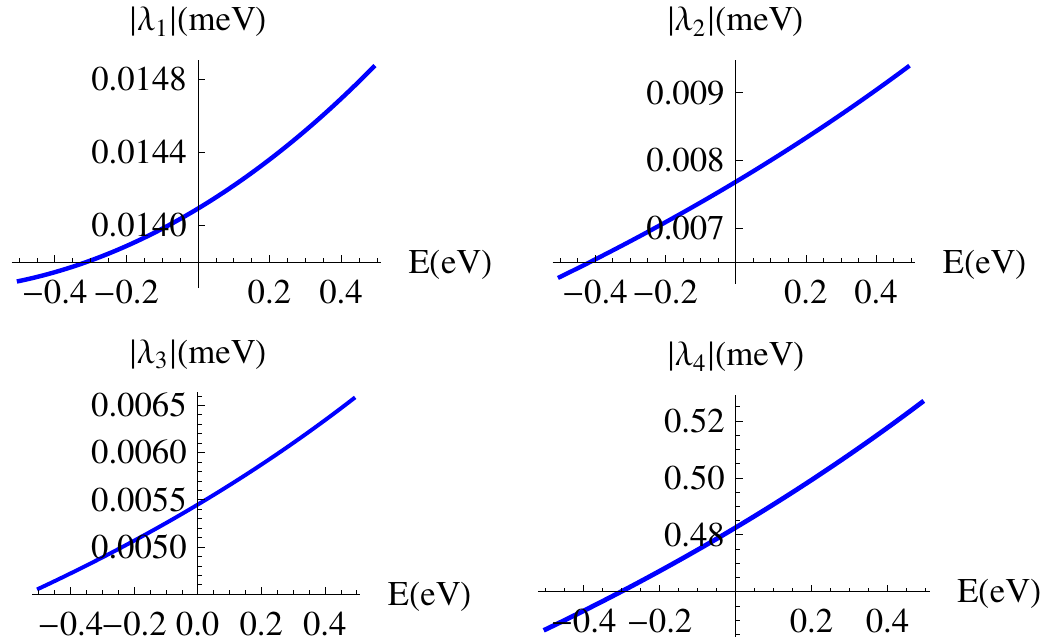}
\caption[fig]{(Color online). Dependence on energy of the spin-orbit
couplings, as defined in eq.~\ref{hamilso}.}
\label{couplings_bilayer_so_E}
\end{figure}

The energy dependence of the four couplings in eq.~\ref{hamilso} is
shown in Fig.~\ref{couplings_bilayer_so_E}. The values of the
couplings scale linearly with $\Delta_{so}$. This dependence can be
understood by treating the next nearest neighbor interlayer coupling
and the intratomic spin-orbit coupling as a perturbation. The
spin-orbit coupling splits the spin up and spin down states of the
$\sigma$ bands in the two layers. The interlayer couplings couple
the $\pi$ band in one layer to the $\sigma$ band in the other layer.
Their value is of order $\gamma_3$.  The $\pi$ states are shifted
by:
\begin{align}
\delta \epsilon_{\pi \pm} &\sim - \frac{\gamma_3^2}{\left|
\epsilon_{\sigma \pm}\right|} \propto \mp \Delta_{so} \left(
\frac{\gamma_3}{ \epsilon_{\sigma}^0 } \right)^2
\label{spin_hopping}
\end{align}
where $\epsilon_\sigma^0$ is an average value of a level in the
$\sigma$ band.

 The model gives for the only
intrinsic spin-orbit coupling in single layer graphene the value
\begin{align}
\left| \lambda_1^{SLG} \right| &= 0.0065 {\rm meV} \label{lambdaSLG}
\end{align}
This coupling depends quadratically on $\Delta_{so}$, $\delta
\epsilon_{\pi \pm} \sim \pm \Delta_{so}^2 /
\epsilon_\sigma^0$\cite{HGB06}.

The band dispersion of bilayer graphene at low energies, in the
absence of spin-orbit couplings is given by four Dirac cones,
because of trigonal warping effects associated with
$\gamma_3$\cite{MF06}. Hence, we must to consider the couplings for
wavevectors $\vec{\bf k}$ slightly away from the $K$ and $K'$
points. We have checked that the dependence of the couplings
$\lambda_i$ on momentum, in the range where trigonal warping is
relevant, is comparable to the changes with energy shown in
Fig.~\ref{couplings_bilayer_so_E}.

\begin{figure}
\includegraphics[width=8cm]{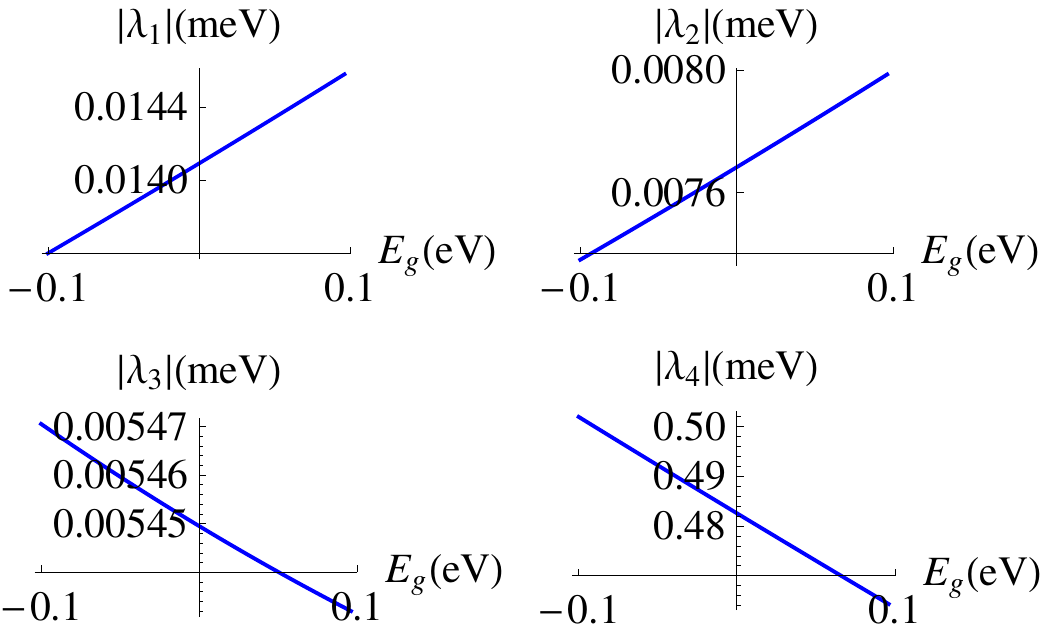}
\caption[fig]{(Color online). Dependence on interlayer gap, $E_g$,
of the spin-orbit couplings, as defined in eq.~\ref{hamilso}.}
\label{couplings_bilayer_so_Egap}
\end{figure}

A gap, $E_g$, between the two layers breaks inversion symmetry, and
can lead to new couplings. The calculations show no new coupling
greater than $10^{-6}$meV for gaps in the range $-0.1 {\rm eV} \le
E_g \le 0.1 {\rm eV}$. The dependence of the couplings on the value
of the gap is shown in Fig.~\ref{couplings_bilayer_so_Egap}. This
calculation considers only the effect in the shift of the
electrostatic potential between the two layers. The existence also
of an electric field will mix the $p_z$ and $s$ orbitals within each
atom, leading to a Rashba term similar to the one induced in single
layer graphene\cite{HGB06,HMING06t}.

The effect of $\lambda_1$ is to open a gap of opposite sign in the
two valleys, for each value of $s_z$. The system will become a
topological insulator\cite{H88,KM05}. The number of edge states is
two, that is, even. The spin Hall conductivity is equal to two
quantum units of conductance. A perturbation which preserves time
reversal invariance can hybridize the edge modes and open a gap.
Such perturbation should be of the form $\tau_x s_y$.

The terms with $\lambda_3$ and $\lambda_4$ describe spin flip
hoppings which involve a site coupled to the other layer by the
parameter $\gamma_1$. The amplitude of the wavefunctions at these
sites is suppressed at low energies\cite{MF06}. The shifts induced
by $\lambda_3$ and $\lambda_4$ in the low energy electronic levels
will be of order $\lambda_3^2 / \gamma_1 , \lambda_4^2 / \gamma_1$.

\begin{figure}
\includegraphics[width=8cm]{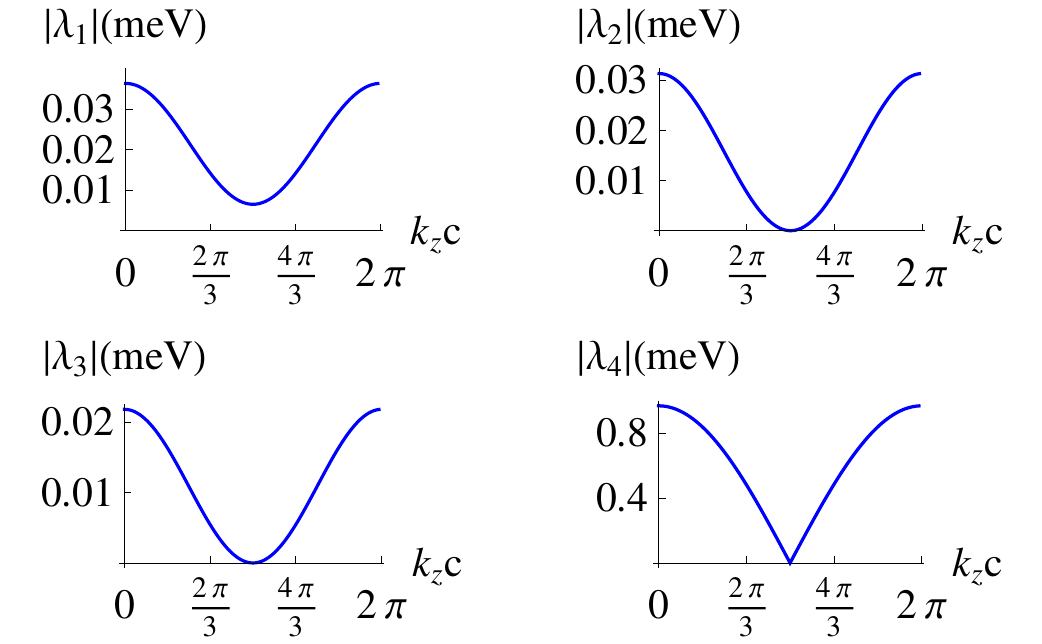}
\caption[fig]{(Color online). Dependence on momentum perpendicular
to the layers in Bernal graphite of the spin-orbit couplings, as
defined in eq.~\ref{hamilso}.} \label{couplings_bilayer_so_kz}
\end{figure}
\subsection{Bulk graphite}.
The hamiltonian of bulk graphite with Bernal stacking can be reduced
to a set of bilayer hamiltonians with interlayer hoppings which
depend on the momentum along the direction perpendicular to the
layers, $k_z$. We neglect in the following the (small) hoppings
which describe hoppings between next nearest neighbor layers,
$\gamma_2$ and $\gamma_5$, and the energy shift $\Delta$ between
atoms in different sublattices. At the $K$ and $K'$ points of the
three dimensional Brillouin Zone ($2 k_z c = 0$, where $c$ is the
interlayer distance) the hamiltonian is that of a single bilayer
where the value of all interlayer hoppings is doubled. At the $H$
and $H'$ points, where $2 k_z c = \pi$, the hamiltonian reduces to
two decoupled layers, and in the intermediate cases the interlayer
couplings are multiplied by $| 2 \cos (k_z c) |$. Carrying out the
calculations described in the previous section, $k_z$ dependent
effective couplings, $\lambda_i ( k_z )$, can be defined. These
couplings are shown in Fig.~\ref{couplings_bilayer_so_kz}. The
results for bilayer graphene correspond to $k_z c = 2 \pi / 3 , 4
\pi / 3$. The layers are decoupled for $k_z c = \pi$. In this case,
the only coupling is $\lambda_1$, which gives the coupling for a
single layer, given in eq.~\ref{lambdaSLG}.

The significant dispersion as function of momentum parallel to the
layers shown in Fig.~\ref{couplings_bilayer_so_kz} implies the
existence of spin dependent hoppings between layers in different
unit cells. This is consistent with the analysis which showed that
the spin-orbit coupling in a bilayer has a contribution from
interlayer hopping, see eq.~\ref{spin_hopping}.

The spin-orbit couplings can be larger in bulk graphite than in a
graphene bilayer. The bands in Bernal graphite do not have
electron-hole symmetry. The shift in the Fermi energy with respect
to the Dirac energy is about $E_F \approx 20 {\rm meV} \gg \lambda_1
, \lambda_3$\cite{DM64}. Hence, the spin-orbit coupling is not
strong enough to open a gap throughout the entire Fermi surface, and
graphite will not become an insulator.

\begin{figure}
\includegraphics[width=8cm]{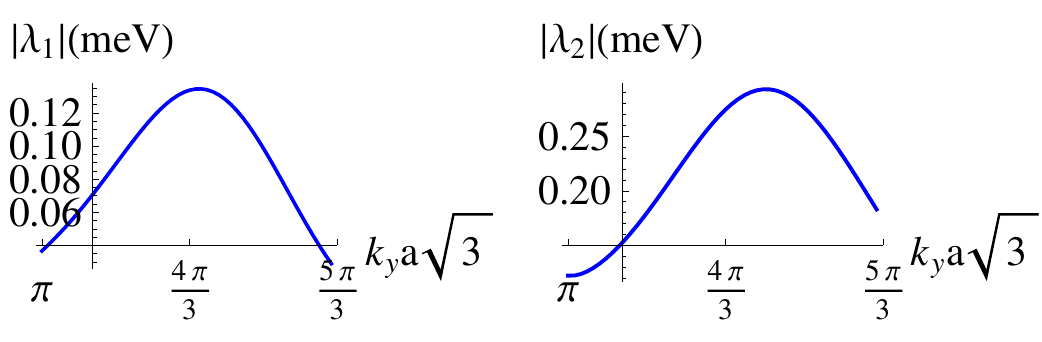}
\caption[fig]{(Color online). Dependence on wavevector, $2 k_y $, of
the spin-orbit couplings for orthorhombic graphite, as defined in
eq.~\ref{couplings_ortho}. The point $k_x = 0 , k_y a \sqrt{3} = 4
\pi /3$ corresponds to the $K$ point ($a$ is the distance between
carbon atoms in the plane).} \label{couplings_bilayer_so_ky_ortho}
\end{figure}

A similar analysis applies to orthorhombic graphite, which is
characterized by the stacking sequence $ABCABC \cdots$\cite{M69}.
The electronic structure of this allotrope at low energies differs
markedly from Bernal graphite\cite{GNP06,AG08}, and it can be a
model for stacking defects\cite{BCP88,GNP06,AG08}. If hoppings
beyond nearest neighbor layers are neglected, the hamiltonian can be
reduced to an effective one layer hamiltonian where all sites are
equivalent. The effective hamiltonian which describes the $K$ and
$K'$ valleys contains eight entries, which can be described using
the matrices $\sigma_i , s_i$, and $\tau_i$. Orthorhombic graphene
is not invariant under inversion, and a Rashba like spin-orbit
coupling is allowed. The spin-orbit coupling takes the form:
\begin{align}
{\cal H}_{ortho}^{so} &\equiv \lambda_1^{ortho} \sigma_z s_z \tau_z
+ \lambda_2^{ortho} \left( \sigma_y s_x - \tau_z \sigma_x s_y
\right) \label{ortho}
\end{align}
As in the case of Bernal stacking, the couplings have a significant
dependence on the momentum perpendicular to the layers, $k_z$, and
interlayer hopping terms are induced. For $\omega = 0, \vec{\bf k} =
0$ and $k_z = 0$, we find:
\begin{align}
\lambda_1^{ortho} &= 0.134 {\rm meV} \nonumber \\
\lambda_2^{ortho} &= 0.275 {\rm meV} \label{couplings_ortho}
\end{align}
In orthorhombic graphite the Fermi level is away from the $K$ and
$K'$ points, in the vicinity of a circle defined by $| \vec{\bf k} |
= \gamma_1 / v_F$\cite{GNP06,AG08}. The variation of the couplings
as function of wavevector is shown in
Fig.~\ref{couplings_bilayer_so_ky_ortho}.

\section{Conclusions}
We have studied the intrinsic spin-orbit interactions in a graphene
bilayer and in graphite. We assume that the origin of the couplings
is the intraatomic $\vec{\bf L} \vec{\bf S}$ interaction, and we use
a tight binding model which includes the $2s$ and $2p$ atomic
orbitals.

The intrinsic spin-orbit couplings in a graphene bilayer and in
graphite are about one order of magnitude larger than in single
layer graphene, due to mixing between the $\pi$ and $\sigma$ bands
by interlayer hoppings. Still, these couplings are typically of
order $0.01 - 0.1$meV, that is, $0.1 - 1$K.

Bilayer graphene becomes an insulator with an even number of edge
states. These states can be mixed by perturbations which do not
break time reversal symmetry. These perturbations can only arise
from local impurities with strong spin-orbit coupling, as a spin
flip process and intervalley scattering are required.

The interplay of spin-orbit coupling and interlayer hopping leads to
spin dependent hopping terms. The spin-orbit interactions are
largest in orthorhombic graphite, which does not have inversion
symmetry.

\section{Acknowledgements}
Funding from MICINN (Spain), through grants FIS2008-00124 and
CONSOLIDER CSD2007-00010 is gratefully acknowledged.
\bibliography{spin_orbit_relaxation_10}
\end{document}